\documentclass[10pt,conference]{IEEEtran}

\usepackage{alltt}
\usepackage{amsmath}
\usepackage{array}
  \newcolumntype{L}[1]{>{\raggedright\let\newline\\\arraybackslash\hspace{0pt}}m{#1}}
  \newcolumntype{C}[1]{>{\centering\let\newline\\\arraybackslash\hspace{0pt}}m{#1}}
  \newcolumntype{R}[1]{>{\raggedleft\let\newline\\\arraybackslash\hspace{0pt}}m{#1}}
\usepackage{arydshln}
\usepackage{balance}
\usepackage[inline]{enumitem}
\usepackage{fancyvrb}
  \fvset{fontsize=\scriptsize,xleftmargin=3mm,frame=leftline,framerule=0.5mm,rulecolor=\color{gray}}
\usepackage[para]{footmisc}
\usepackage{framed}
\usepackage{graphicx}
  \graphicspath{{assets/figures/}}
  \DeclareGraphicsExtensions{.pdf,.eps,.png}
\usepackage[utf8]{inputenc}
\usepackage{lipsum}
\usepackage{multirow}
\usepackage{pmboxdraw}
\usepackage{pifont}
\usepackage{rotating}
\usepackage[normalem]{ulem}
\usepackage[hyphens,spaces,obeyspaces]{url}
  \makeatletter
  \g@addto@macro{\UrlBreaks}{\UrlOrds}
  \makeatother
\usepackage{hyperref}
\usepackage{cleveref}
 \crefformat{section}{\S#2#1#3}
  \crefformat{subsection}{\S#2#1#3}
  \crefformat{subsubsection}{\S#2#1#3}
\usepackage{xcolor}
\usepackage{textcomp}
\usepackage{graphicx}
\usepackage{caption} 
\usepackage{dblfloatfix}    
\usepackage{comment}
\usepackage{multirow}

\newcommand{\ttt}[1]{\texttt{#1}}						            

\newcommand{\blind}[1]{\textcolor{gray}{#1}}               

\begin{document}

\title{ChestyBot: Detecting and Disrupting Chinese Communist Party Influence Stratagems}


\author{
  \blind{Matthew Stoffolano}, \blind{Ayush Rout}, \blind{Justin M. Pelletier} \\
    Rochester Institute of Technology\\
    Email: \ttt{jxpics@rit.edu}\\
    \\

\textit{“We’re surrounded. That simplifies the problem.”} -Chesty Puller    
\vspace{-1.5em}

}

\maketitle

\begin{abstract}
Foreign information operations conducted by Russian and Chinese actors exploit the United States' permissive information environment. These campaigns threaten democratic institutions and the broader Westphalian model. Yet, existing detection and mitigation strategies often fail to identify active information campaigns in real time. This paper introduces \textit{ChestyBot}, a pragmatics-based language model that detects unlabeled foreign malign influence tweets with up to 98.34\% accuracy. The model supports a novel framework to disrupt foreign influence operations in their formative stages.

\end{abstract}



\section{Introduction}\label{sec:introduction}

Foreign influence campaigns—particularly those attributed to Russia during the 2016 U.S. Presidential Election—demonstrated how state-sponsored social media operations can destabilize democratic societies \cite{badawy2019characterizing}. During that campaign, social media posts emanating from one state -- Russia -- probably represented an intentional effort to influence the internal affairs of another country -- the United States. Though these efforts may not have changed election outcomes, they nonetheless constitute an erosion of the Westphalian state model itself \cite{lukito2020coordinating}. 

In recent years, China has attempted to use social media to influence foreign perceptions of internal matters such as the Beijing 2022 Winter Olympics, the origins of COVID-19, and the human rights abuses in Xinjiang  \cite{jacobs2022whodefinesdemocracy}. Despite these initiatives, China has (as far as we can tell at the time of this writing) not performed a successful large-scale disinformation campaign directed against U.S. internal interests. However, in recent years China's attacks have adopted similar mannerisms to Russia \cite{repnikova2022people}, which could increase the risk of impact.  

Standard ways to detect influence campaigns include looking for bot networks \cite{pacheco2020uncovering}, observing coordinating behavior \cite{vargas2020detection}, and implementing various types of retrospective analysis \cite{carley2020social}. Though useful in retrospect, most methods struggle to detect campaigns as they unfold. Early detection of foreign malign influence is important because taking preventive measures in the early stages of an influence campaign on social media can permanently stunt the campaign's growth \cite{lewandowsky2021countering}. Improving the current methods requires looking at a combination of the attacker methods and victim communities. 

To better understand the trajectory of this threat and the potential for foreign malign influence to impact U.S. internal affairs, this project aims to provide a proof of concept for detecting the influence campaigns undertaken by the Chinese Communist Party(CCP)/People's Liberation Army(PLA) in its earliest stages. We, therefore, propose a novel method to detect four influence stratagems used by the CCP, with data generated during the 2022 U.S. Mid-term Elections.

\section{Background}\label{sec:Background}

\subsection{Social Cybersecurity}\label{socialcyber}
Social science seeks to understand how innate human traits affect the world \cite{hinde1987individuals}. In recent years, researchers have leveraged computational social science, societal computing, data science, and policy studies in the emergent field called Social Cybersecurity \cite{national2019decadal}. This field considers linguistic, business, group, and behavioral analyses using techniques such as dating mining or forensics \cite{carley2020social}.

\subsection{Information Warfare}\label{infowar}
Information Warfare manages the narratives that reach a population to achieve decisive ideological outcomes. This can be done by dropping pamphlets over a specific region or spreading disinformation on social media \cite{farwell_2021}. The strategies used against the United States are visualized below in Figure \ref{information}.

\begin{figure*}[!t]
  \centering
  \includegraphics[width=\textwidth]{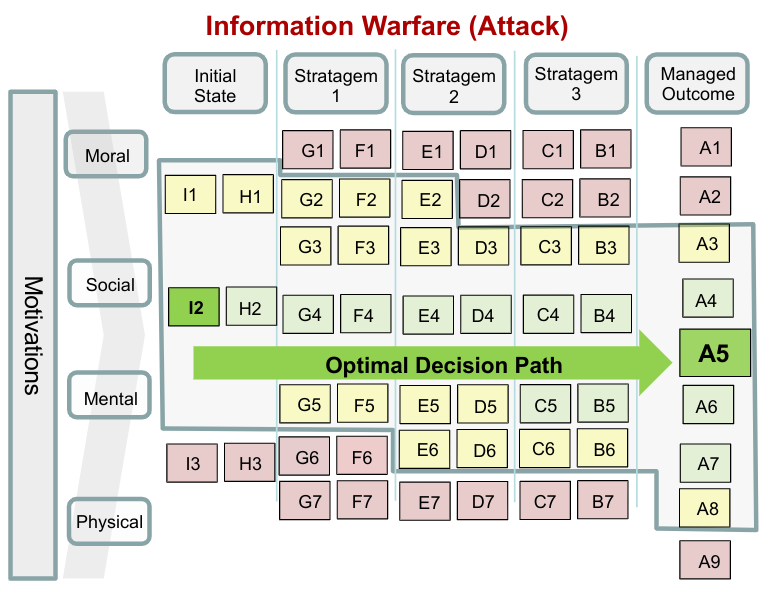}
  \caption{Each colored box represents a decision that can be influenced by the application of Stratagems. The introduction of stimuli prompts the actor to make a decision beneficial to an influencer’s selected course of action [Starting with decision I2]. The deliberate application of Stratagems directs decisions toward a specific outcome [Green Arrow to A5] and, failing that, accounts for the range of possible decisions that would lead to acceptable managed outcomes [Green Boxes]. Adapted from communications with Colonel Drew Cukor, USMC in 2019.}
  \label{information}
\end{figure*}

\subsection{Influence Stratagems}\label{stratagems}
Influence Campaigns consist of multiple operations employing various tactical-level strategies (called \textit{stratagems} throughout this paper for clarity). These stratagems are probably best described by The Taxonomy of Influence Strategies, provided as an artifact of a NATO-funded examination of Russia's annexation of the Crimean Peninsula \cite{kelly_paul_2020}, depicted in Figure \ref{TaxonomyPeriodicTable}.

Though Russia is more stentorian than China in foreign malign influence directed against the United States, we stipulate that there are likely overlaps between the Russian propaganda playbook and what China is likely to use. We discuss this more in section \ref{sec:Method}.



\subsection{Echo Chambers}\label{echochambers}
Implementing algorithms on social media to improve the user experience has created a phenomenon known as \textit{echo chambers}. Members of these isolated communities interact mostly or exclusively with other community members, with few or no outside influences. Mapping these communities shows how information flows in the online fora they frequent. Figure \ref{echo} depicts the flow of information between two chambers through an intermediary (a liminal node), which represents high betweenness centrality. This measure of influence is an important component of identifying relationships among echo chambers in social networks \cite{kratzke2023find}.

\section{Related Work}

Several attempts to analyze and detect disinformation exist. It is beyond the scope of this paper to provide an exhaustive listing. One important exemplar of this body of knowledge, conducted by researchers at Lincoln Labs, focuses on finding disinformation and identifying critical actors. Their methodology focuses on starting with a keyword search to build a narrative \cite{smith2021automatic}. This in part informed our approach to look for keywords that could indicate an influence stratagem in the datasets that aligned with previous methods used by the CCP to validate ChestyBot's training model.

Each nation, however, does have variation in how they conduct influence campaigns.  During the Covid pandemic, the CCP employed operations to spread their narrative by trying to redirect through sharing positive stories and recast through rewriting history.\cite{molter2020pandemics} Further, in further events such as the Beijing protest, the CCP showed more strategies by seeking to inform the global population with information meant to affect the public's view of the events.  They would also attempt to deflect events by framing the protesters in morally compromising positions, such as spreading stories of them using M320 a tool used in the United States military for employing High Explosives.  All of these narratives are pushed with the idea of spreading the Chinese stories.\cite{diresta2020telling}

Also, the analysis of echo chambers has been performed by many other researchers seeking to understand how misinformation spreads. The most common focus reveals indicators of a bad actor releasing misinformation with the intent of others picking it up and spreading it. Sometimes this is done in conjunction with social media bots that amplify the misinformation \cite{diaz2023disinformation}. Though our approach is similar, we extend this pursuit by looking for detection of influence campaigns, which can contain misinformation, disinformation, and even true information that is framed in a specific way.

\begin{figure}[ht!]
    \centering
    \includegraphics[width=\columnwidth]{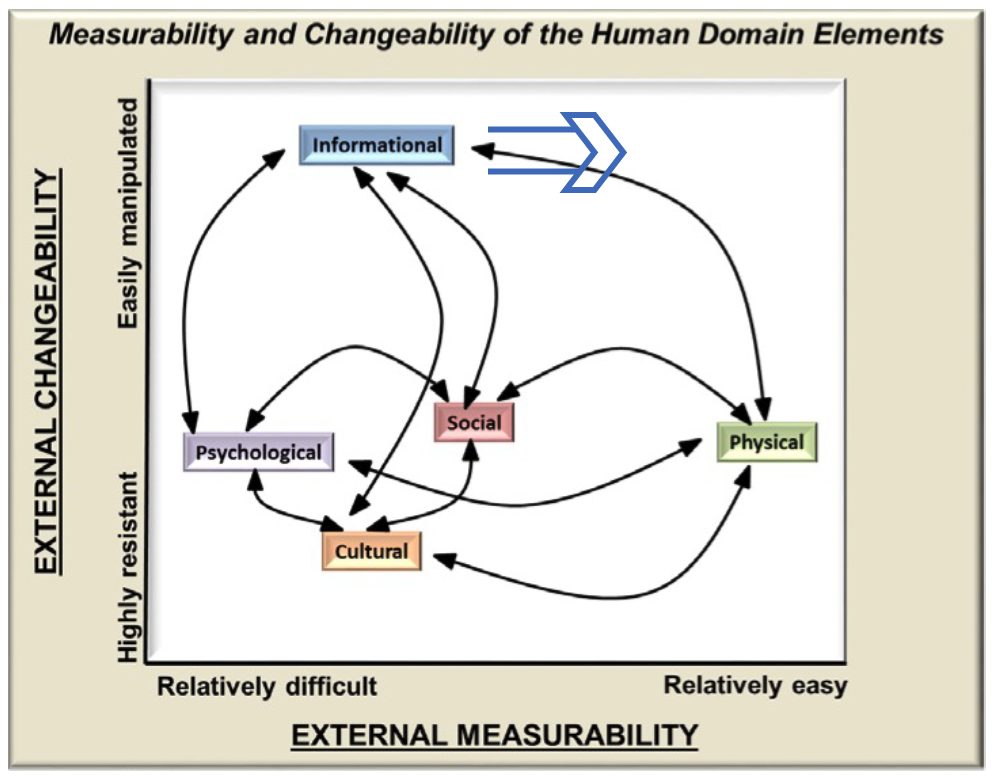}
     \caption{Our main effort is to make the external measurability of informational artifacts easier, which is a problem well described in a 2015 U.S. Special Operations Command publication \cite{command2015operating}.}
     \label{approach}
\end{figure}

Finally, a 2015 U.S. Special Operations Command publication titled "Operating in the Human Domain" specifically highlighted the difficulty of measuring information \cite{command2015operating}. The same publication also featured the ease of manipulating information as a fundamental challenge of achieving desired affects. This observation in particular informs our approach to frame the problem we seek to solve as one that improves the \textit{measurability} of information operations. This approach is depicted in Figure \ref{approach}.

\section{Method}\label{sec:Method}







We employed the snowball method, starting with accounts with known ties to the CCP and adding identified aligned accounts (layer one) as well as another layer of accounts (layer two) who are aligned with layer one accounts.  We consider this snowball to be a representative sample of an echo chamber.  We also employed the Louvain Algorithm to identify echo chambers among a wider set of data for testing purposes. After this sampling, we employed a natural language process to detect possible propaganda tweets according to the four specific influence strategies (stratagems): inform, invoke, deflect, and recast. The remainder of this section describes each component of this method in detail.

\subsection{Platform Selection}
Data was collected on X (Twitter) over other platforms because of the data type, access, and tools.   Any platform focused on images or videos was excluded to remove complexity and focus on text communication.  Text-based communication was further narrowed down based on site allowance to give researchers access to data and a tool that could implement with the collection\cite{chaudhary2021twitter}.  Combined with Tweepy\cite{roesslein2009tweepy} and research access to X (Twitter), we were able to pull down current user information to build data sets for training and testing.

\subsection{Snowball Sampling Method (SSM)}
Discovering any group that is trying to employ an influence campaign is met with an inherent problem due to the clandestine nature of the operation. Thus, it became necessary to employ the snowball sampling method(SSM) to collect the necessary data to research operatives employing these campaigns. SSM emerged as a useful method for finding hard-to-reach groups\cite{goodman1961snowball, parker2019snowball} and is used by anthropologists to locate groups in a conflict environment where inherent mistrust exists\cite{cohen2011field}. Further, SSM has also been implemented on social media to find political dynamics in unitary countries\cite{clark2006field}. SSM is ideal for characterization of influence behaviors emanating from known or suspected influencers because it has been widely used in behavioral research\cite{yoshida2013snowball, noy2008sampling} and remains one of the most popular methods of sampling in qualitative research that examine characteristics of networking and referral\cite{parker2019snowball, naderifar2017snowball}.

SSM makes it possible to create a network containing accounts suspected of involvement in an influence campaign. Each echo chamber shows user account interactions and the proliferation of ideas across the echo chamber. In this case, SSM began with identifying known CCP accounts and performing a recursive search on friends, followers, and retweets.  In aggregate, this allowed us to build a list of potential accounts that could be involved in an influence campaign emanating from one actor "Jane Doe". Using the results of SSM, we constructed a data set of potential influence activities, (\texttt{Set1}), described below.

\begin{figure}[ht!]
        \centering
        \includegraphics[width=0.8\columnwidth]{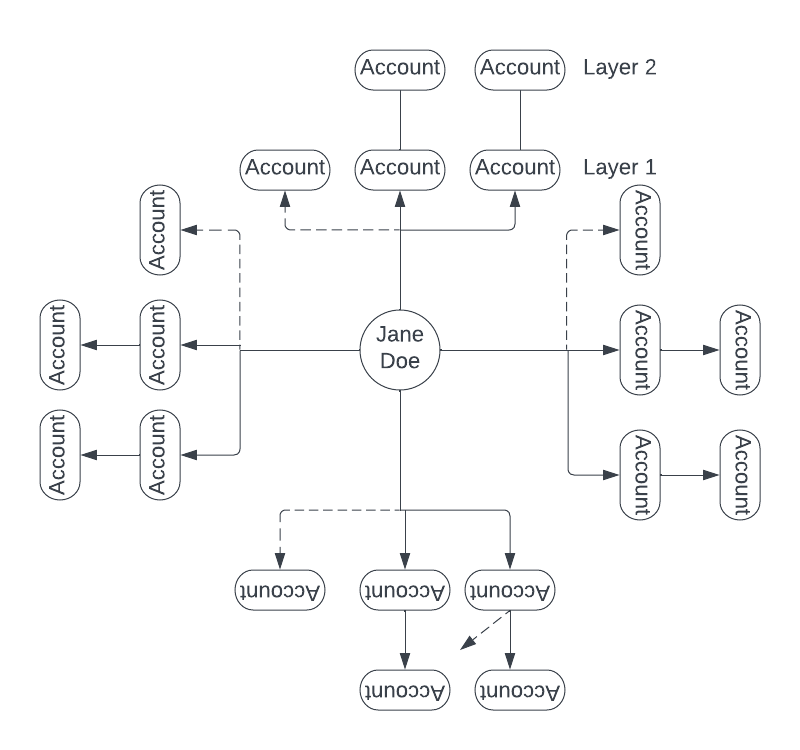}
         \caption{We used the Snowball Sampling Method to build \texttt{Set1}.}
         \label{snowwball}
    \end{figure}

\subsection{Community Identification and Edge Weighting}
The \textbf{Louvain Algorithm} was implemented on \texttt{Set2} to map out the echo chambers. The Algorithm originated in 2008 and has emerged as the de-facto standard for quickly extracting community structure from large networks \cite{blondel2008fast}. The Louvain Algorithm is a heuristic method based on modularity optimization that analyzes large web graphs to identify communities. It works in two phases: (1) modularity optimization through local community changes and (2) aggregation of communities to build a new network of communities. These phases repeat until no modularity gain is possible. We utilized the Louvain Algorithm because of its ability to map out vast online communities in a short period of time, as well as its ability to identify the artificial nodes  \cite{kido2016topic}, and the influential nodes in a network \cite{joshi2023identifying}.  This allowed for quickly assembling the echo chambers and being able to understand the role of nodes in the community.

We optimized the communities by defining \textbf{edge weights}, where \textit{likes} received a weight of one and \textit{retweets} and \textit{friends/followers} received a weight of 10. 

We optimized the communities by defining \textbf{edge weights}, where \textit{likes} received a weight of one and \textit{retweets} and \textit{friends/followers} received a weight of 10.

\subsection{Data Retrieval and Anonymization}
Two datasets were collected (\texttt{Set1} and \texttt{Set2}) using different methods based on the SSM.  Both \texttt{Sets} contained a hash of the user ID, creation date, and tweet contents.

\texttt{Set1} started with Jane Doe, a CCP member who commonly pushes influences narratives.  Pulling her tweets from 1 October 2022 to 8 November 2022 provided the information to construct the first layer using SSM. For each of Jane’s tweets, we collected 20 unique random users who retweeted the tweet. The new accounts comprise the first layer, and then the same process of tweet search is conducted on the recently collected accounts to form the second and final layers. In total, there were approximately 10,000 tweets collected, a majority of those were posted in the English language. This process is depicted in summary in Figure \ref{snowwball}.

\texttt{Set2} started with a list of CCP members, including Jane Doe, and extended to retrieve data on friends/followers, retweets, and likes. Using the Louvain Algorithm, we assigned each account to an echo chamber, as depicted in Figure \ref{echo}. Then one of the echo chambers was randomly selected to collect tweets from 1 January 2010 to 25 February 2023. This resulted in a dataset of approximately 280,000 tweets.

    \begin{figure*}[ht!]
        \centering
        \includegraphics[width=\textwidth]{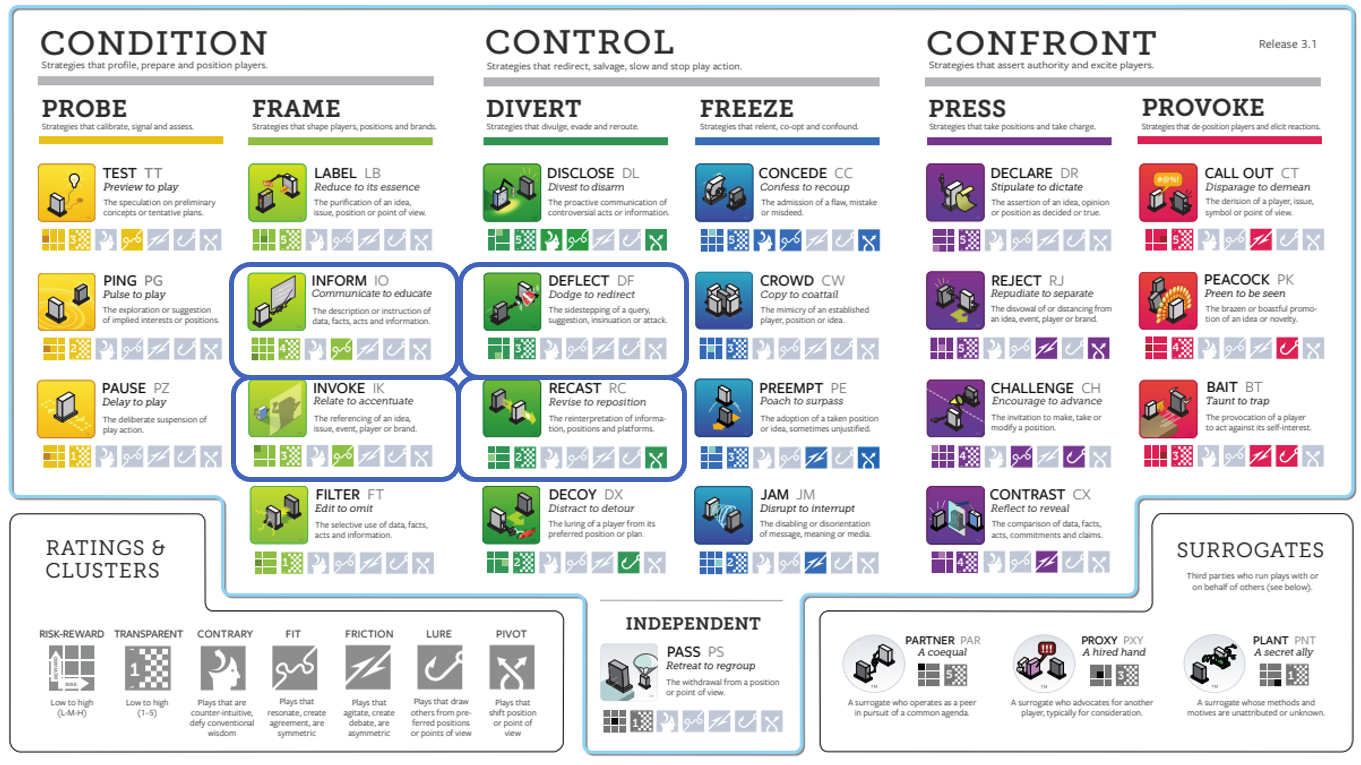}
         \caption{The NATO-funded taxonomy \cite{kelly_paul_2020} depicted here provides a digest of Russian influence stratagems used during the annexation of Crimea.} 
         \label{TaxonomyPeriodicTable}
    \end{figure*}
\subsection{Data Labeling}\label{labeling}
Both \texttt{Set1} and \texttt{Set2} data sets initially contained a hashed user ID, the creation date, and a tweet.  \texttt{Set1} was used for training and therefore encoded according to four influence stratagems: inform, invoke, deflect, and recast. These influence strategems were identified and described in a NATO-funded taxonomy\cite{kelly_paul_2020} Though Russia is probably more likely to use a different range of influence stratagems than China, we believe that Chinese Communist Party members are also likely to use the four highlighted stratagems (\textbf{inform}, \textbf{invoke}, \textbf{deflect}, \textbf{recast}) from the Russian propaganda playbook.

\texttt{Set2} was used for testing. Encoding was done manually where each tweet was labeled according to the presence of any of the four stratagems.  If we found the tweet matched one of the stratagems, it would be marked as true.  A tweet only needed to have one true strategem to become classified as a tweet that was part of an influence campaign.

Each of these sets, and the influence stratagems themselves, are defined in the remainder of this sub-section. The entire taxonomy of influence stratagems and our highlighted selections are depicted below in Figure \ref{TaxonomyPeriodicTable}.
\\

\noindent \texttt{Set1}\\
\noindent Inform and invoke focus on framing the debate.

\textbf{Inform} focuses on educating users about specific information. One prominent example we found is the Nordstrom Pipeline Leak, as in "\#ChinaDailyCartoon Who gains most from \#NordStream sabotage?"

\textbf{Invoke} focuses on connecting to a particular event. One of the invoked tweets used in training is "US shames MS, while ignores its own vile human rights records -George Floyd...".

\hspace{1em}

\noindent Deflect and recast are diversion tactics, where attackers attempt to spin a conversation. 

A common form of deflection is bringing up the US treatment of minorities during international human rights discussions.  An example \textbf{deflect} tactic we observed in \texttt{Set1} stated: "Firearm, abortion, economy, immigration ... all become ammunition for U.S. partisan fight, a fight making the house of American politics totter". 

Recasting is simply another term for rebranding. For example, an example \textbf{recast} tactic we observed in the snowball stated: "NATO is not a defense alliance, it's a war machine. Ask the people of Afghanistan, Iraq or Libya".
\\

If the tweet was determined to have one of these four stratagems, it was marked as a propaganda tweet. If not, we encoded it as not a propaganda tweet. This led to 882 total tweets encoded, with 62 of those designated as propaganda.
\\

\noindent \texttt{Set2} \\
\texttt{Set2} was encoded using dates and keywords; each set of dates was a time that a campaign had been expected to have occurred. Though useful in concept, we found the list of keywords to be vastly inferior as an indicator of campaign following the building and testing of ChestyBot. 

\subsection{Building ChestyBot}

Through classifying small data using known influence stratagems, a detector-bot using machine learning can detect potential tweets of a campaign. Once one tweet is detected, a defender can further inspect the area where the campaign occurs. Improving early detection and tracking campaign contagion provides essential tools for studying orchestrated influence campaigns. This is a critical task for our effort.

There exist promising models that might help automate the identification of influence stratagems within tweets, as in Tensor Flow's natural language processing(NLP) toolset\cite{ganegedara2018natural}. In particular, we considered two pre-trained models (\texttt{ROBERTO} and \texttt{BERT}) but found them incompatible. These tools and many others like them have an over-emphasis on sentiment analysis in text prediction but do not explicitly consider the logical rationale or context employed within the language \cite{koroteev2021bert, yue2019survey}. The logical rationale and context -- what someone means -- is called \textit{pragmatics}, while the emotion behind the words -- how someone feels -- is called \textit{semantics}. Our observation of this limitation among available models largely conforms to previous findings of Li, Thomas, and Liu, who call on academic researchers and NLP developers to place less emphasis on semantics and more on pragmatics \cite{li2021semantics}. Therefore, instead of using pre-trained models, we trained a pragmatics-focused NLP model (ChestyBot) using \texttt{Set1} and tested it using \texttt{Set2}.  



We chose a convolutional neural network (CNN) for this task of text classification, rather than a traditional feedforward neural network, because CNNs are better able to capture the local patterns and dependencies in the input text data \cite{gu2018recent}. Text data exhibits a strong sequential dependency, where the meaning of a word is shaped by the context in which it appears. Traditional feedforward neural network models, which treat input data as a fixed-size vector, fail to fully account for these contextual dependencies. Specifically, the feedforward network cannot distinguish between the different senses of a given the word when used in various contexts. On the other hand, a CNN is able to capture the local patterns in the input data through the use of convolutional filters. In the context of text classification, the filters can be designed to recognize common n-grams (i.e., contiguous sequences of n words) in the input text data. The pooling layers can then be used to downsample the output of the convolutional layers, which reduces the dimensionality of the input data while preserving the most salient features \cite{gholamalinezhad2020pooling}. In addition to capturing local patterns, a CNN can also learn hierarchical representations of the input data. Each layer in the CNN learns progressively more complex representations of the input data, allowing the model to capture both low-level features such as individual words, as well as higher-level features such as phrases or sentences. 

For the purposes of this investigation, we built a CNN architecture using TensorFlow's Keras library. The input data consisted of a set of text strings from \texttt{Set1}, which were encoded as one-hot vectors of length 'input\_dim', where input\_dim was set to 1536 which represents the size of the vocabulary used to encode the text. One hot encoding is a way of numerically representing text where each word in the vocabulary is assigned a unique integer index \cite{hancock2020survey}. Each element of the vector is set to zero, except for the element corresponding to the index of the word in the vocabulary, which is set to one. The resulting data is then padded to a fixed length of input\_length using the tf.keras.preprocessing.sequence.pad\_sequences() function. The encoded and padded data is then passed through an embedding layer, which maps each index to a dense vector representation of length dense\_vectors. 

The output of the embedding layer is a 3D tensor of shape (num\_examples, input\_length, dense\_vectors), where each element corresponds to the embedding of a word. Following the embedding layer, a 1D convolutional layer with 32 filters of size 5 is applied to the embeddings. This layer performs a convolution operation along the temporal dimension of the input, capturing local patterns in the sequence of embeddings. The output of this layer is a 3D tensor of shape (num\_examples, new\_input\_length, num\_filters), where new\_input\_length = input\_length - 4 due to the size of the filters used. Subsequently, a 1D max pooling layer is applied to the output of the convolutional layer, which reduces the dimensionality of the output by selecting the maximum value over a window of size 2. The resulting output is a 3D tensor of shape (num\_examples, new\_input\_length/2, num\_filters). At this point, the output of the max pooling layer is flattened using the tf.keras.layers.Flatten() layer. This operation reshapes the output into a 2D tensor of shape (num\_examples, new\_input\_length/2 * num\_filters), where each row corresponds to an example in the batch and each column corresponds to a flattened feature. Finally, the flattened output is passed through a dense layer with a single output unit and a sigmoid activation function, which produces a binary classification output. In simple terms, this classification output was "propaganda" or "not propaganda", based on the influence stratagem encodings in \texttt{Set1}.

\begin{figure}[ht!]
        \centering
        \includegraphics[width=0.9\columnwidth]{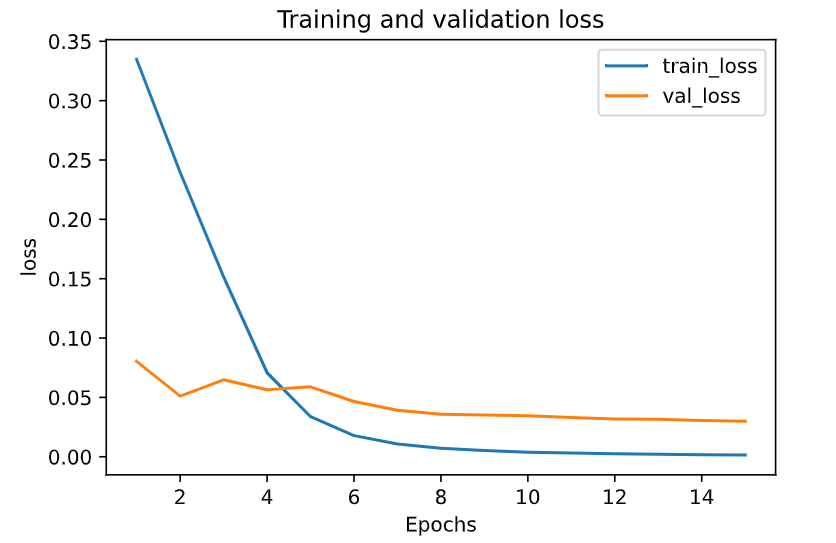}
         \caption{Training for ChestyBot}
         \label{parser}
    \end{figure}

Ultimately, this led to the creation of ChestyBot, which revealed efficient training (depicted in Figure \ref{parser}) and promising test performance (described in the next section). 

\section{Findings}\label{Findings}
Our research focused on finding methods to detect a campaign in the early stages when the impact is still containable.  Using SSM, we found it is possible to discover previously undetected accounts that were part of a campaign. Also, by using data generated by these accounts, we were able to map communities and determine how each one was connected. Together, this revealed the liminal nodes that may share information across other echo chambers. We were also able to apply a novel pragmatics-based NLP model (ChestyBot) that was able detect a campaign in one of these chambers. The remainder of this section describes more fully our proof of concept model.

ChestyBot performed well in identifying influencing tweets; during training, it had a validation accuracy of 99.44 percent. Testing it against a random echo chamber in \texttt{Set2}, ChestyBot found 241 potential propaganda tweets. 92.53 percent of those were True Positive (Obvious True), 5.81 percent of tweets were accurately classified according to the context (Context-Dependent True), and 1.66 percent false positives. Of note, though it would benefit from fine-tuning to increase the likely hood of detecting tweets, fine-tuning to catch \textit{all} propaganda tweets is probably unnecessary because our objective is reliable detection of ongoing campaigns rather than every part of every campaign. Even so, ChestyBot could be improved through additional data labeling and training, which would allow for improved indicators and warning of a campaign underway. These improvements are likely to minimize false positives that occur from an abstract connection.
\\

\noindent \textit{Obvious True:}
\begin{itemize}
  \item "Tariffs are not the right way to go. American businesses and consumers want solutions, not sanctions"
    \item "It's not just your local supermarket. Eggs are 60 percent more expensive than last year in the U.S"
  \item "The issues related to Xinjiang \& Hong Kong are not about democracy, human rights, ethnicity or religion, but about anti-terrorism, anti-secession, about protecting people's lives, safeguarding China's national sovereignty, security, and development interests". 
  \item "China and the US do compete. But how should the competition play out? Our relations should not be like the intensely confrontational American football match. There should be no offensive team or defensive team, no touchdown, no quarterback sack."
  \item"59 people have been killed in 36 mass shootings in the U.S. since the start of the year." 
\end{itemize}
\vspace{3mm} 
\noindent \textit{Context-Dependent True:}
  \begin{itemize}
    \item "How a country is doing on human rights is essentially gauged by whether the interests of its people are upheld, and whether they enjoy a growing sense of fulfillment, happiness and security"
  \end{itemize}
  \vspace{3mm} 
  \noindent \textit{False Positive:}
  \begin{itemize}
    \item "The United Nations report on the PRC’s human rights violations against the Uyghur people and other minority groups in Xinjiang is clear and devastating.  The U.S. calls on Beijing to cease these atrocities"
    \item “Everyone thought the billionaires could save us — clearly, they're not engaged,” grumbled one executive at a billionaire-owned newsroom. “It's not a good model because they lose interest or they get pissy.” 
  \end{itemize}

\section{A Novel Framework for Influence Operation Disruption}\label{Disruption}

The automated detection of malign influence campaigns enables a more proactive approach to social cybersecurity, particularly when combined with network analysis techniques. Once an influence campaign is detected using tools such as ChestyBot, a key step in disrupting its efficacy is to identify the \textit{liminal nodes}—users or accounts that bridge otherwise distinct echo chambers.

    \begin{figure}[ht!]
        \centering
        \includegraphics[width=0.8\columnwidth]{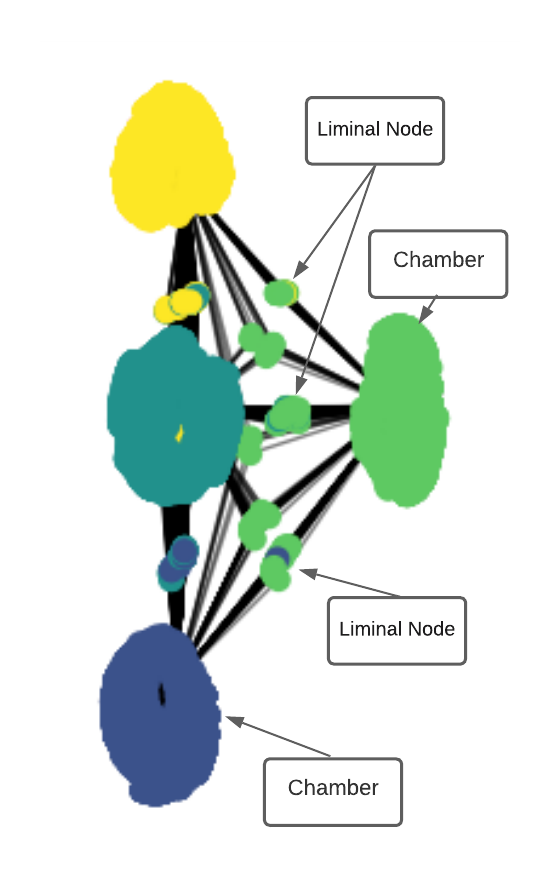}
         \caption{We incorporated the Louvain Algorithm to build \textit{echo chambers} and identify \textit{liminal nodes} between multiple \textit{chambers}.}
         \label{echo}
    \end{figure}

These liminal nodes exhibit high betweenness centrality, meaning they occupy structurally important positions in the social graph that allow them to serve as information conduits between insular communities. Messages originating from a malign actor and reaching these nodes are more likely to ``break out'' of their original echo chamber and achieve broader dissemination.

By coupling early detection of propaganda tweets with echo chamber community mapping (e.g., via the Louvain Algorithm), we can observe where influence operations are gaining traction and predict their breakout potential. When a malign message is picked up by a liminal node, that message becomes a candidate for targeted disruption—such as counter-messaging, deprioritization in feeds, or account moderation.

Operationally, this means that automated detection systems should not only flag malicious content but also evaluate the network position of users who engage with it. A tweet that reaches a highly central liminal node may pose more strategic risk than one that circulates exclusively within a single community.

This framework transforms detection from a passive, forensic activity into an active defense mechanism capable of shaping the informational battlespace in real time. As depicted in Figure \ref{echo}, the integration of message classification and network topology offers a dynamic framework for mitigating malign influence at the point of maximum potential impact.

\section{Conclusions and Discussion}\label{Discussion}
The connection of social media has created a relatively new domain for information warfare. Around the world, many countries are deploying varying strategies to win the hearts and minds of foreign populations. However, many strategies in this domain remain underdeveloped possibly due to an over-focus on technology and retrospective analysis. Our approach focuses on how information flows in the human domain, and we suggest that engaging with these technologies can facilitate a better understanding of and defense against influence campaigns. 

The Snowball Sampling Method allows for finding accounts that may participate in a campaign and mapping the communities around these accounts. This allowed for the deployment of ChestyBot, which can detect potential campaign tweets based on the stratagems that an attacker employs. The ability to detect these tweets during a live campaign allows information that would otherwise remain hidden from other researchers. This can allow for early campaign detection. 

Furthermore, the Louvain Algorithm allows the isolation of specific communities and the identification of liminal nodes connecting echo chambers. Each of these liminal nodes is a weak point in an adversary's campaign; it could represent a sort of \textit{key terrain}. We propose that facilitating effects against those nodes probably provides a way to manage information flow between online communities.

By combining these methods, we were able to provide a proof of concept mechanism -- ChestyBot -- capable of detecting influence campaigns emanating from the Chinese Communist Party. Furthermore, we propose that such a detection mechanism can inform follow-on effects (ie. block, canalize, contain, defeat, destroy, disrupt, fix, interdict, isolate, neutralize, suppress, turn) on foreign malign influence campaigns. We believe our proof of concept and the effects it allows could help re-gain the initiative in the information sphere.

\section{Limitations and Future Work}
Though X (Twitter) provides many advantages, retrospective analyses are limited due to limitations on the information that is accessible/collected. This impacted our ability to build a complete scenario. For example, we could not see when one account became another's friend/follower, so the communities collected may have accounts that were community members during the campaign. Furthermore, our access to seeing \textit{likes} was limited to only the most recent 100. Even though we believe the effects from this temporal bias were minimized by prioritizing edge weights for \textit{retweets} and \textit{following} activities, we believe a more robust real-time data collection could improve echo chamber construction.

Two recommended advances for  ChestyBot are to (1) improve accuracy and (2) to generalize detection capabilities. To improve accuracy, we point out that \texttt{Set1} data only contained 62 tweets labeled as propaganda. Though ChestyBot performed well (better than expected, really), a more substantial data set would almost certainly improve the ability to detect different forms of the trained influence stratagems.  Further, ChestyBot was only trained on 4 stratagems that would match likely Chinese methods. Including more data and exploring ChestyBot's capability to detect other stratagems would make it more useful in the field. Thogether, these improvements would probably make our proof of concept model better able to detect foreign malign influence campaigns emanating from other actors across the competition continuum.

Finally, there remain important and necessary ethical considerations regarding the weaponization of ChestyBot and the associated influence operation disruption framework. Many major systems will have something to say about this. We consider only a few here as fodder for future inquiry. From a Catholic ethical standpoint, the inherent dignity of every human person demands caution when deploying technologies that may surveil or manipulate information environments, even in the name of defense\cite{compendium2004}. Classical liberal democratic norms similarly require that interventions in public discourse respect individual autonomy, freedom of expression, and the pluralism essential to a free society\cite{mill1859, rawls1999}. Yet from a military operational perspective, the growing sophistication of adversarial influence operations necessitates proactive capabilities that can detect and neutralize emerging threats in the information domain. These tensions reveal a critical tradeoff: the very tools that safeguard national sovereignty and civilian morale may, if ungoverned or misapplied, erode the moral and civic foundations they aim to protect. Therefore, any operational use of such systems must be accompanied by rigorous oversight, strict proportionality, clear attribution, and a reaffirmed commitment to both just war principles and the preservation of open democratic dialogue.

\section{Acknowledgements}
We would like to acknowledge the support and contributions from CPT Noah Demoes and entire the team at CyberRecon. J.M. Pelletier would like to express deep gratitude for the ongoing support he receives from the \emph{Ordo Praedicatorum}.



\bibliographystyle{IEEEtran}
\bibliography{bibfile}




\end{document}